\documentclass[seceq]{ptptex}

\usepackage{wrapft}
\usepackage{graphicx} 
\usepackage{amssymb}
\usepackage{amsmath}

\newcommand{\Slash}[1]{{\ooalign{\hfil/\hfil\crcr$#1$}}}

\title{
Direct Detection of Vector Dark Matter
}

\author{
Junji \textsc{Hisano}$^{1,2}$,
Koji \textsc{Ishiwata}$^{3}$
Natsumi \textsc{Nagata}$^{1,4}$ 
and Masato \textsc{Yamanaka}$^{5}$
}

\inst{
$^1$Department of Physics, Nagoya University, Nagoya 464-8602, Japan\\
$^2$Institute for the Physics and Mathematics of the Universe,
University of Tokyo, Kashiwa 277-8568, Japan \\
$^3$California Institute of Technology, Pasadena, CA 91125, USA\\
$^4$Department of Physics, University of Tokyo, Tokyo 113-0033, Japan\\
$^5$MISC, Kyoto Sangyo University, Kyoto 603-8555, Japan
}

\abst{
  In this paper, we complete formulae for the elastic 
  scattering cross section of
  general vector dark matter with nucleons in the direct detection at
  the leading order of the strong coupling constant $\alpha_s$,
  assuming that the dark matter is composed of vector particles and
  interacts with heavy fermions with color charge as well as
  standard-model quark.  As an application of our formulae, the direct
  detection of the first Kaluza-Klein photon in the minimal universal
  extra dimension model is discussed.  It is found that the 
  scattering cross
  section is larger than those in the previous works by up to a factor
  of ten.  }

\begin{document}
\maketitle

\section{Introduction}

\label{sec:intro}

Cosmological observations have established the existence of dark
matter. Nowadays, the energy density of dark matter is precisely
determined using the Wilkinson Microwave Anisotropy Probe (WMAP) 
satellite \cite{Hinshaw:2008kr}.  
The standard model of particle physics, however, can explain neither 
its existence nor its nature, which has been a mystery in particle 
physics and cosmology.  The Weakly Interacting Massive Particles 
(WIMPs) in models beyond the standard model are a good 
candidate for dark matter. 
The relic abundance is naturally consistent with the cosmological 
observation if WIMPs have TeV-scale mass and they are thermally 
produced in the early universe. This is the so-called thermal relic 
scenario. In this scenario, the dark matter is originally nonrelativistic 
and acts as cold dark matter in the era of the structure formation 
of the universe.  The leading candidate of WIMP dark matter is the 
lightest neutralino in the minimal supersymmetric standard model, 
which is a Majorana fermion. 
In addition, other models beyond the standard model at the TeV scale 
predict the existence of stable vector particles, such as the
Kaluza-Klein (KK) photon \cite{Servant:2002aq,Cheng:2002ej} in the
universal extra dimensions (UED) \cite{Appelquist:2000nn,Cheng:2002iz}
and the $T$-odd heavy photon \cite
{Hubisz:2004ft,Birkedal:2006fz,asano} in the Littlest Higgs model with
$T$-parity \cite{ArkaniHamed:2001nc,Cheng:2004yc}.

Now, the WIMP dark matter scenario has been tested in collider and
direct detection experiments.  In the Large Hadron Collider (LHC),
(pair) production of the WIMP dark matter with TeV-scale mass is
expected.  On the other hand, XENON100 \cite{Aprile:2011hi}, which is
the largest-volume detector ever, is in operation to detect it as 
scattering signals with nuclei on the earth.  If the WIMP-scattering
events are discovered and its properties are measured in the direct
detection experiments, the measurements would be tested using the 
data that indicates dark matter production at the LHC.

With such recent progress on the experimental side, the theoretical
prediction of dark matter signals must have better accuracy.  The
collider signatures of dark matter (such as in supersymmetric or
UED models) have been intensively studied.  On the other hand, the
scattering cross section of dark matter with nuclei in the direct
detection experiments has also been calculated more accurately. In 
recent works \cite{Hisano:2010fy,Hisano:2010ct}, it was pointed out
that a gluon-WIMP effective interaction is one of the leading
contributors to the elastic scattering cross section, and it was 
also shown how to evaluate it precisely. (In those works, the results 
are shown for supersymmetric dark matter.) 
For the other models, in contrast, the gluon contribution and other 
leading terms are not taken into account correctly in the calculation of 
the scattering cross section (\cite{Cheng:2002ej, Servant:2002hb, 
Arrenberg:2008wy} for the UED model and \cite{Birkedal:2006fz} for 
the Littlest Higgs model with $T$-parity).

In this paper, we assume that WIMP dark matter is composed of vector
particles, and evaluate its elastic scattering cross section with 
nucleons, which is relevant to the direct detection experiments. Here, 
we consider the case in which vector dark matter interacts with 
standard-model quarks and exotic heavy fermions, which is a 
fundamental representation of $SU(3)_c$. As an application of our 
formulae, we discuss the direct detection of the first KK photon in the 
minimal universal extra dimension (MUED) model.  We found that the 
spin-independent scattering cross section with a proton is 
increased by up to a factor of ten, compared with those in the previous 
works.  We also show that the spin-independent scattering cross 
section of the KK photon with a proton ranges from about $3 \times 
10^{-46} \text{cm}^2$ to $5 \times 10^{-48} \text{cm}^2$ on the 
parameter region with the thermal relic abundance consistent with the 
cosmological observations.  Future direct detection experiments with 
ton-scale detectors might cover the range.

This paper is organized as follows. In the next section, we briefly
review the formulation for the scattering cross section of general vector
dark matter with nuclei. Next, in \S\ref{sec:EffLag}, we derive the
effective Lagrangian describing the interaction of general vector dark
matter with quarks and gluons in target nuclei.  In \S\ref{KK}, we
review the MUED model and summarize the mass spectra of KK particles,
and apply the formulae that are generally derived to the case of the
first KK photon dark matter. Then, we numerically calculate the 
scattering cross section of the first KK photon dark matter 
with/without higher KK mode contributions, and discuss the significance 
of each effective operator.  Section \ref{sec:conc} is devoted to the 
conclusion.

\section{Formalism of cross section for vector dark matter}
\label{sec:eff}

In this section, we formulate the elastic cross section of general
vector dark matter with nuclei.  The cross section is calculated in
terms of effective coupling constants, which are given by coefficients
of effective interactions of vector dark matter with light quarks
($q=u,d,s$) and gluons.  To begin with, we write down all the effective
interactions that are relevant to the scattering.  In the derivation, 
we referred to reference.~\cite{Drees:1993bu}.

Since the scattering process is nonrelativistic, all the terms that
depend on the velocities of dark matter particles and nuclei are
subdominant in the velocity expansion. Therefore, we neglect the
operators suppressed by the velocities of the dark matter particles or
nuclei in our study.  Then, physical degrees of freedom of the vector
field, which we denote as $B_{\mu}$ with the mass $M$, where we also
assume a real field, are restricted to their spatial components since
$\partial_{\mu}B^{\mu}=0$.  As a consequence, in the expansion of
a strong coupling constant and in the nonrelativistic limit of the
scattering process, the leading interaction of the vector field with
quarks and gluons is given as
\begin{equation}
\mathcal{L}^{\mathrm{eff}}=\sum_{q=u,d,s}\mathcal{L}^{\mathrm{eff}}_q
+\mathcal{L}^{\mathrm{eff}}_G, 
\end{equation}
with
\begin{eqnarray}
  \mathcal{L}^{\mathrm{eff}}_q &=&
  f^m_q m_q B^{\mu}B_{\mu}\bar{q}q
  +f^{i{\tiny \Slash{D}}}_q  B^{\mu}B_{\mu}\bar{q}i\Slash{D}q+
\frac{d_q}{M}
  \epsilon_{\mu\nu\rho\sigma}B^{\mu}i\partial^{\nu}B^{\rho}
  \bar{q}\gamma^{\sigma}\gamma^{5}q+\frac{g_q}{M^2}
  B^{\rho}i\partial^{\mu}i\partial^{\nu}B_{\rho}\mathcal{O}^q_{\mu\nu},
\nonumber \\ 
 \label{eff_lagq}
\\
\mathcal{L}^{\mathrm{eff}}_G&=&f_G
 B^{\rho}B_{\rho}G^{a\mu\nu}G^a_{\mu\nu}
 +\frac{g_G}{M^2}B^{\rho}i\partial^{\mu}i\partial^{\nu}B_{\rho}
 \mathcal{O}^g_{\mu\nu}  
 \label{eff_lagG},
\end{eqnarray} 
where $m_q$ is the mass of a quark and $\epsilon^{\mu\nu\rho\sigma}$ is
the totally antisymmetric tensor defined as $\epsilon^{0123}=+1$.
Here, the covariant derivative is defined as $D_\mu\equiv\partial_\mu+i
g_sA^a_\mu T_a$, with $g_s$, $T_a$, and $A^a_\mu$ being the $SU(3)_C$
coupling constant, generator, and the gluon field,
respectively. $\mathcal{O}^q_{\mu\nu}$ and $\mathcal{O}^g_{\mu\nu}$
are the twist-2 operators (traceless parts of the energy-momentum
tensor) for quarks and gluons, respectively,
 \begin{eqnarray} 
{\cal O}_{\mu\nu}^q&\equiv&\frac12
\bar{q} i \left(D_{\mu}\gamma_{\nu} + D_{\nu}\gamma_{\mu}
  -\frac{1}{2}g_{\mu\nu}\Slash{D} \right) q \ ,
\\
{\cal O}_{\mu\nu}^g&\equiv&\left(G_{\mu}^{a\rho}G_{\rho\nu}^{a}+
  \frac{1}{4}g_{\mu\nu} G^a_{\alpha\beta}G^{a\alpha\beta}\right) \ .
\label{twist2}
\end{eqnarray}
Here, $G^a_{\mu \nu}$ is the field strength tensor of a gluon.  In the
derivation of the effective Lagrangian, we used the equations of
motion for vector field, {\it i.e.}, $(\partial^2-M^2)B_{\mu}=0$.  The
third term in the right-hand side of Eq.\ (\ref{eff_lagq}) contributes
to the spin-dependent interaction in the scattering of dark matter
with nuclei, while the other terms in the effective Lagrangian
contribute to the spin-independent one.

Scattering amplitude is given by the matrix element of the effective
interaction put between initial and final states. In this evaluation,
we use the equation of motion for light quarks. The validity of the
application is proven in the case where the operator is evaluated using 
on-shell hadron states~\cite{Politzer:1980me}. Then, denoting
$|N\rangle$ ($N=p,n$) as the on-shell nucleon state, we can evaluate the
second term in Eq.\ (\ref{eff_lagq}) as
\begin{eqnarray}
\langle N| \bar{q} i \Slash{D} q |N \rangle =
\langle N|m_q \bar{q}  q |N \rangle .
\end{eqnarray}
Now, let us first examine the spin-independent scattering. As in the
scattering process, the momentum transfer is negligible; the matrix
elements between initial and final nucleon states with the mass
$m_N~(N=p,n)$ are obtained as
\begin{eqnarray}
 \langle N \vert m_q \bar{q} q \vert N\rangle/m_N  &\equiv& f_{Tq}\ ,
\nonumber
\\
 1-\sum_{u,d,s}f_{Tq} &\equiv& f_{TG}  \ ,
\nonumber\\
\langle N(p)\vert 
{\cal O}_{\mu\nu}^q
\vert N(p) \rangle 
&=&\frac{1}{m_N}
(p_{\mu}p_{\nu}-\frac{1}{4}m^2_N g_{\mu\nu})\
(q(2)+\bar{q}(2)) \ ,
\nonumber\\
\langle N(p) \vert 
{\cal O}_{\mu\nu}^g
\vert N(p) \rangle
& =& \frac{1}{m_N}
(p_{\mu}p_{\nu}-\frac{1}{4}m^2_N g_{\mu\nu})\ 
G(2) \ .
\end{eqnarray}
In the matrix elements of twist-2 operators, $q(2)$, $\bar{q}(2)$, and
$G(2)$ are the second moments of the parton distribution functions
(PDFs) of a quark $q(x)$, antiquark $\bar{q}(x)$ and gluon $g(x)$,
respectively,
\begin{eqnarray}
q(2)+ \bar{q}(2) &=&\int^{1}_{0} dx ~x~ [q(x)+\bar{q}(x)] \ ,
\cr
G(2) &=&\int^{1}_{0} dx ~x ~g(x) \ .
\end{eqnarray}
Then, after the trivial calculation of matrix elements for vector dark
matter states, we obtain a spin-independent effective coupling of vector
dark matter with nucleons, $f_N$, as
\begin{eqnarray}
f_N/m_N&=&\sum_{q=u,d,s}
f_q f_{Tq}
+\sum_{q=u,d,s,c,b}
\frac{3}{4} \left(q(2)+\bar{q}(2)\right)g_q
\nonumber\\
&-&\frac{8\pi}{9\alpha_s}f_{TG} f_G 
+\frac{3}{4} G(2)g_G \ , 
\label{f}\end{eqnarray}
where $f_q\equiv f^m_q+f^{i{\tiny \Slash{D}}}_q$ and $\alpha_s =
g_s^2/4 \pi$.  Note that the effective scalar coupling of dark matter
with gluons, $f_G$, gives the leading contribution to the cross section
even if it is suppressed by a one-loop factor compared with those of
light quarks. On the other hand, the contributions with the twist-2
operators of gluons are of the next-leading order as $g_G$ is 
suppressed by $\alpha_s$.  Thus, we ignore them in this paper.

The spin-dependent scattering can be dealt with in a similar manner; the 
spin-dependent effective coupling is given by
\begin{eqnarray}
a_{N}&=&\sum_{q=u,d,s} d_q \Delta q_N \ ,
\end{eqnarray}
with
\begin{eqnarray}
2 s_{\mu}\Delta q_N &\equiv& \langle N \vert 
\bar{q}\gamma_{\mu}\gamma_5  q \vert N \rangle \ ,
\end{eqnarray}
where $s_{\mu}$ is the spin of a nucleon.

\begin{table}
  \caption{ Parameters for quark and gluon matrix elements used in this paper. 
    $f_{Tq}$ are evaluated using the results in 
    Refs.~\cite{Corsetti:2000yq, Ohki:2008ff, Cheng:1988im}. The spin 
    fractions for a proton are obtained from  Ref.~\cite{Adams:1995ufa}. 
    The second moments of PDFs of gluons, quarks, and antiquarks in 
    a proton are calculated at the scale $\mu=m_Z$ using 
    the CTEQ parton distribution \cite{Pumplin:2002vw}.} 
\label{table1}
\begin{center}
\begin{tabular}{ccc}
\begin{minipage}{0.25\hsize}
\begin{center}
\begin{tabular}{|l|l|}
\hline
\multicolumn{2}{|c|}{For proton}\cr
\hline
$f_{Tu}$& 0.023\cr
$f_{Td}$& 0.032\cr
$f_{Ts}$&0.020\cr
\hline
\multicolumn{2}{|c|}{For neutron}\cr
\hline
$f_{Tu}$&0.017\cr
$f_{Td}$& 0.041\cr
$f_{Ts}$& 0.020 \cr
\hline
\end{tabular}
\end{center}
\end{minipage}
\begin{minipage}{0.25\hsize}
\begin{center}
\begin{tabular}{|l|l|}
\hline
\multicolumn{2}{|c|}{Spin fraction }\cr
\multicolumn{2}{|r|}{(for proton)}\cr
\hline
$\Delta u_p$& 0.77\cr
$\Delta d_p$& -0.49\cr
$\Delta s_p$& -0.15\cr
\hline 
\end{tabular}
\end{center}
\end{minipage}
\begin{minipage}{0.5\hsize}
\begin{center}
\begin{tabular}{|l|l||l|l|}
\hline
\multicolumn{4}{|c|}{Second moment at $\mu=m_Z$ }\cr
\multicolumn{4}{|r|}{(for proton) }\cr
\hline
$G(2)$&0.48&&\cr
$u(2)$&0.22&$\bar{u}(2)$& 0.034\cr
$d(2)$&0.11&$\bar{d}(2)$&0.036\cr
$s(2)$&0.026&$\bar{s}(2)$&0.026\cr
$c(2)$&0.019&$\bar{c}(2)$&0.019\cr
$b(2)$&0.012&$\bar{b}(2)$&0.012\cr
\hline
\end{tabular}
\end{center}
\end{minipage}
\end{tabular}
\end{center}

\end{table}

For our calculation, we describe, in Table~\ref{table1}, the input
parameters for the hadronic matrix elements that we use in this
article.  The mass fractions and spin fractions (in a proton) of light
quarks, $f_{Tq}$ and $\Delta q_p$, are evaluated using the results in
Refs.~\cite{Corsetti:2000yq, Ohki:2008ff, Cheng:1988im} and
Ref.~\cite{Adams:1995ufa}, respectively.  The prescription for
evaluating the mass fraction is described in these references and
Ref.~\cite{Hisano:2010ct}.

The second moments of PDFs of gluons, quarks, and antiquarks in
a proton, on the other hand, have scale dependence.  
We use the values that are calculated at the scale $\mu=m_Z$ ($m_Z$ 
is $Z$ boson mass). This is because all the terms (and coefficients) in
effective couplings should be evaluated at the scale of the mass of
the particle that mediates the scattering.\footnote
 {
   Here, some readers may concern about it by considering $1/Q^2$ ($Q$
   is momentum transfer) expansion in the operator-product expansion
   method, such as in the evaluation of the deep inelastic scattering
   cross section of hadrons. However, that is not what we are
   considering. Rather, we derive the effective Lagrangian in a similar
   manner to when one derives the weak-interaction Lagrangian at low
   energy from a fundamental theory. When one gives a theoretical
   prediction of physical observables from the weak-interaction
   effective Lagrangian, one should set a renormalization scale in the
   evaluation of the matrix element of the physical amplitude to the
   weak scale, {\it i.e.,} the mass scale of the particle that mediates
   the scattering, in order to make the theoretical prediction
   accurate without calculating higher order quantum corrections.
   This is why PDFs at the weak scale are a good approximation.
}
It is obvious at tree level scattering. Even in the case where the loop 
diagram induces scattering, it is understandable from the fact that the 
loop integration is dominated by the particle mass.\footnote
  {We thank Mark B. Wise for pointing this out.}
The numerical values of the second moments of PDFs are
shown in Table~\ref{table1} using the CTEQ parton distribution
\cite{Pumplin:2002vw}.  The spin fractions and second moments for
a neutron are given by the exchange of up and 
down quarks.

Eventually, using $f_N$ and $a_N$, we express the elastic scattering
cross section of vector dark matter with target nuclei as follows:
\begin{equation}
 \sigma=
\frac{1}{\pi}\biggl(\frac{m_A}{M+m_A}\biggr)^2\biggl[|n_pf_p+n_nf_n|^2
+\frac{8}{3}\frac{J+1}{J}|a_p\langle S_p\rangle
+a_n\langle S_n\rangle|^2\biggr].
\end{equation}
Here, $m_A$ is the mass of target nucleus $A$.  $n_p$ and $n_n$ are
proton and neutron numbers in the target nucleus, respectively, while
$J$ is the total spin of the target nucleus and $\langle S_N\rangle=
\langle A\vert S_N\vert A\rangle$ are the expectation values of the
total spin of protons and neutrons in $A$.  The first term in the
brackets on the right-hand side comes from the spin-independent
interactions, while the second one is generated by the spin-dependent
one. 

\section{Effective Lagrangian}
\label{sec:EffLag}

In this section, we derive the effective Lagrangian for general vector
dark matter, which was introduced in the previous section. As we
described in the Introduction, we assume that vector dark matter
has interactions with fermions with color charge, in which quarks are
included. Denoting  these fermions as $\psi_1$ and $\psi_2$ with 
masses $m_1$ and $m_2$, respectively, we start with a simply 
parametrized interaction Lagrangian of the form:
\begin{eqnarray}
  \mathcal{L}=
  \bar{\psi}_2 ~(a_{\psi_2\psi_1} \gamma^{\mu}+b_{\psi_2\psi_1}
  \gamma^{\mu}\gamma_5)\psi_1  B_{\mu}   + \mathrm{h.c.}
\label{eq:simpleL}
\end{eqnarray}
Here, we do not consider the case in which vector dark matter couples
to a pair of quarks, since otherwise it would become unstable.  We take
$m_2>m_1$ without loss of generality here and thereafter.  We also
assume, for simplicity, that $\psi_1$ and $\psi_2$ are color
triplets. It is straightforward to extend to the other cases.

\begin{figure}[t]
 \begin{center}
  \includegraphics[height=4cm]{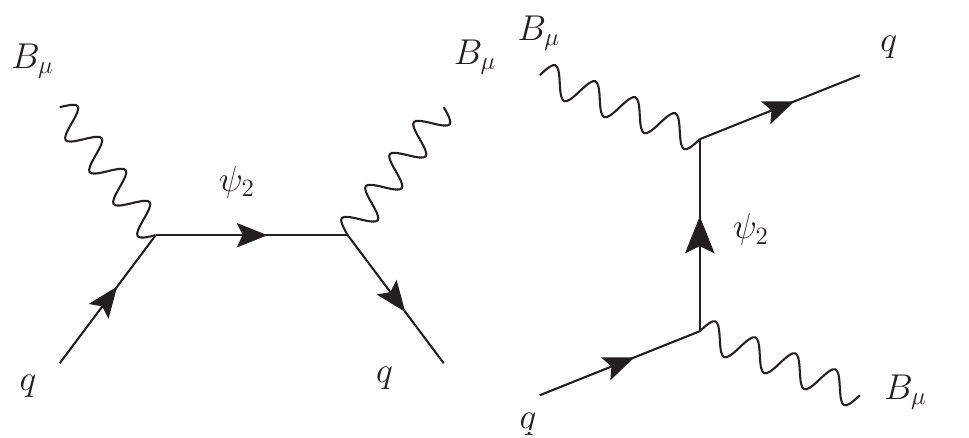}
  \caption{Tree level diagrams of color triplet fermion $\psi_2$
    exchange to generate interaction of vector dark matter with
    quarks.}
\label{fig:tree_general}
 \end{center}
\end{figure}

First, let us examine the scattering with quarks. The vector dark
matter is scattered by quarks at the tree level. The relevant interaction
Lagrangian is given by taking $\psi_1=q$ in Eq.~(\ref{eq:simpleL}),
and the diagrams are shown in Fig.~\ref{fig:tree_general}.  After
integrating out the heavy particle $\psi_2$, the coefficients in
Eq.~(\ref{eff_lagq}), {\it i.e.}, scalar-, axial-vector-, and
twist-2-type couplings with quarks, are induced as follows:
\begin{eqnarray}
f_q&=&\frac{a^2_{\psi_2 q}-b^2_{\psi_2 q}}{m_q}\frac{m_{2}}{m^2_{2}-M^2}
-(a^2_{\psi_2 q}+b^2_{\psi_2 q})
\frac{m^2_{2}}{2(m^2_{2}-M^2)^2},   
\label{tree_general_1}  \\
d_q&=&\frac{iM(a^2_{\psi_2 q}+b^2_{\psi_2 q})}{m^2_{2}-M^2},  
\label{tree_general_2}  \\
g_q&=&-\frac{2M^2(a^2_{\psi_2 q}+b^2_{\psi_2 q})}{(m^2_{2}-M^2)^2}.
\label{tree_general_3}
\end{eqnarray}
Here, we take the zero quark mass limit. As is obvious, the effective
couplings obtained here are enhanced when the vector dark matter and
the heavy colored fermion are degenerate in mass.

\begin{figure}[t]
 \begin{center}
  \includegraphics[height=4cm]{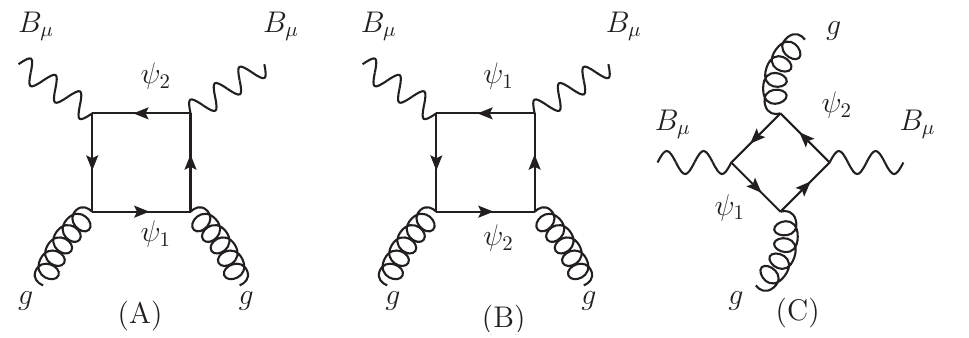}
  \caption{One-loop contributions to scalar-type effective
    coupling with gluons. }
\label{fig:loop_general}
 \end{center}
\end{figure}

Next, let us discuss the interaction with gluons.  The vector dark
matter does not couple with gluons at the tree level; however, it does at
the loop level, as shown in Fig.~\ref{fig:loop_general}.  In those
loop level processes, all the heavy particles $\psi_1$ and $\psi_2$
that couple with $B_{\mu}$ contribute, in addition to quarks.
The contribution of each diagram, (A), (B), and (C), to the
scalar-type effective coupling with gluons $f_G$ in
Eq.~(\ref{eff_lagG}) is written as
\begin{equation}
f_G^{({\rm I})}=
\frac{\alpha_s}{4\pi}
\biggl[(a_{\psi_2 \psi_1}^2+b_{\psi_2 \psi_1}^2)f^{\rm (I)}_{+}(M;m_1,m_2)
+(a_{\psi_2 \psi_1}^2-b_{\psi_2 \psi_1}^2)f^{\rm (I)}_{-}(M;m_1,m_2)
\biggr],
\end{equation}
where $f^{\rm (I)}_{+}$ and $f^{\rm (I)}_{-}$ (I$=$A, B, and C) are
given after the loop calculation,
\begin{multline}
  f^{\rm (A)}_{+}(M;m_1,m_2)=
\frac1{6\Delta^2M^2}
\biggl[\Delta[M^2(m^2_2-m^2_1)+m_1^2(m_1^2+5m_2^2)]\\
-6m_1^2m_2^2[(m_2^2-m_1^2)^2-M^2(m_1^2+3m_2^2)]\biggr] 
  -\frac{m_1^2}{12M^4}\ln\biggl(\frac{m_1^2}{m_2^2}\biggr) \\ 
  +\frac{m_1^2L}{12\Delta^2M^4}\biggl[(m_2^2+m_1^2-M^2)\Delta^2
  +2m_2^2\Delta\{5m_2^4+20m_1^2m_2^2-m_1^4+M^2(9m_2^2+m_1^2)\} \\
  +12m_2^4\{M^2(m_2^4+10m_1^2m_2^2+5m_1^4)-(m_2^2-m_1^2)^2(m_2^2+3m_1^2)\}
\biggr], 
\end{multline}
\begin{multline}
 f^{\rm (A)}_-(M;m_1,m_2)=-\frac{m_2}{6m_1\Delta^2}
\biggl[\Delta(2m_2^2+m_1^2-2M^2)+6m_1^2m_2^2(m_2^2-m_1^2-M^2)\biggr] \\
+\frac{m_1m_2^3\{\Delta+m_1^2(m_2^2-m_1^2+M^2)\}}{\Delta^2}L,
\end{multline}
\begin{multline}
f^{\rm (C)}_+(M;m_1,m_2)=-\frac{2M^4-3(m_1^2+m_2^2)M^2+(m_1^2-m_2^2)^2}
{6\Delta M^2} \\
+\frac{m_1^2}{12M^4}\ln\biggl(\frac{m_1^2}{m_2^2}\biggr)+\frac{m_2^2}
{12M^4}\ln\biggl(\frac{m_2^2}{m_1^2}\biggr) \\
+\frac{\Delta(M^2-m_1^2-m_2^2)(m_1^2+m_2^2)+4m_1^2m_2^2\{(m_1^2-m_2^2)^2
-2M^2(m_1^2+m_2^2)\}
 }{12\Delta M^4}L, 
\end{multline}
and
\begin{eqnarray}
 f^{\rm (B)}_{+}(M;m_1,m_2)&=&f^{\rm (A)}_+(M;m_2,m_1), \\
 f^{\rm (B)}_-(M;m_1,m_2)&=&f^{\rm (A)}_-(M;m_2,m_1), \\
 f^{\rm (C)}_-(M;m_1,m_2)&=&0. 
\label{loopf}
\end{eqnarray}
Here, we defined functions $\Delta$ and $L$ as
\begin{eqnarray}
 \Delta(M;m_1,m_2)&\equiv& M^4-2M^2(m_1^2+m_2^2)+(m_2^2-m_1^2)^2, \\
 L(M;m_1,m_2)&\equiv& \begin{cases}
	   \frac{1}{\sqrt{|\Delta|}}\ln
\biggl(\frac{m_2^2+m_1^2-M^2+\sqrt{|\Delta|}}
{m_2^2+m_1^2-M^2-\sqrt{|\Delta|}}\biggr)&(\Delta
	  >0) \\
\frac{2}{\sqrt{|\Delta|}}\tan^{-1}\biggl
(\frac{\sqrt{|\Delta|}}{m_2^2+m_1^2-M^2}\biggr)&(\Delta<0)
	  \end{cases}.
\end{eqnarray}

The total scalar-type effective coupling with gluons is obtained by the
summation of $f^{\rm (I)}$ for I $=$ A, B and C. Here, one should be
careful in dealing with the contribution of the diagrams in the case where
$\psi_1$ is a quark. In such a case, there exist diagrams in which
momentum with the scale of quark mass dominates the loop integral. We
call them ``long-distance'' contributions. On the other hand, we
describe the contributions from the diagrams in which the loop
momentum with the scale of particle heavier than quarks dominates the
integral as ``short-distance'' ones. When one derives the effective
coupling in the three-flavour effective theory from the six-flavour full
one, one should not include the contribution of light quarks in the
long-distance diagrams, as discussed in
Refs.~\cite{Hisano:2010fy,Hisano:2010ct}.  To identify the
long-distance contribution among diagrams, let us consider the case
where $\psi_1$ is a quark ({\it i.e.}, $m_q\ll m_2$).  In this limit,
the long-distance contribution in $f_G$ should approach to
$-\frac{\alpha_s}{12\pi}f_q$, which is obtained by the calculation of
triangle diagrams for emitting two gluons using the scalar-type effective
coupling with quarks in Eq.~(\ref{tree_general_1})
\cite{Shifman:1978zn};
\begin{eqnarray}
f_{+}^{(A)}(M;m_q,m_2)&\rightarrow&\frac{m_2^2}{6(m_2^2-M^2)^2},  \\
 f_{-}^{(A)}(M;m_q,m_2)&\rightarrow&-\frac{m_2}{3m_q(m_2^2-M^2)},  \\
 f_{+}^{(B)}(M;m_q,m_2)&\rightarrow&\frac{m_2^2}{6(m_2^2-M^2)M^2}
+\frac{m_2^2}{6M^4}\ln\biggl(1-\frac{M^2}{m_2^2}\biggr), \label{eq:fplblimit}\\
 f_{-}^{(B)}(M;m_q,m_2)&\rightarrow&-\frac{m_q(m_2^2-2M^2)}{6m_2(m_2^2-M^2)^2}, \\
 f_{+}^{(C)}(M;m_q,m_2)&\rightarrow&\frac{1}{6(m_2^2-M^2)}-\frac{1}{6M^2}
-\frac{m_2^2}{6M^4}\ln\biggl(1-\frac{M^2}{m_2^2}\biggr).
 \label{eq:fplclimit}
\end{eqnarray}
It is found that diagram (A) gives the long-distance contribution,
whereas the others yield the short-distance ones.  Therefore,
$f_G$ should be given as
\begin{eqnarray}
f_G = 
 \sum_{\psi_2, q=c,b,t}c_q f^{\rm (A)}_G
+ \sum_{\psi_2,q={\rm all}}(f^{\rm (B)}_G+f^{\rm (C)}_G).
\end{eqnarray} 
Here, $c_q=1+11\alpha_s(m_q)/4\pi$ is the additional long-distance QCD
correction pointed out in Ref. \cite{Djouadi:2000ck}.  We evaluate it as
$c_c=1.32$ and $c_b=1.19$ for $\alpha_s(m_Z)=0.118$, and also adopt 
$c_t=1$ for simplicity.  On the other hand, if there are no quarks in
the loop, all the diagrams indicate as short-distance contributions. In such
a case, $f_G$ is simply obtained as
\begin{eqnarray}
f_G = 
 \sum_{\psi_1,\psi_2}(f^{\rm (A)}_G+f^{\rm (B)}_G+f^{\rm (C)}_G).
\end{eqnarray}

\section{Cross section of Kaluza-Klein dark matter with nucleons}
\label{KK}

In this section, we apply the above results to the evaluation for the direct
detection of the lightest Kaluza-Klein particle (LKP) in a model with
UED~\cite{Appelquist:2000nn, Cheng:2002iz}.  Among the variety of UED
models, the MUED model, in which one extra dimension is compactified
on an $S^1/Z_2$ orbifold and a dark matter candidate is included, is
most extensively investigated \cite{Servant:2002aq,Cheng:2002ej}.  In
the MUED model, the stabilization of LKP is ensured by the KK parity
conservation. It originates from the momentum conservation along the
fifth-dimensional space. The KK parity $-1$ ($+1$) is assigned to
particles with odd (even) KK number. This assignment and the KK parity
conservation prevent the lightest KK-odd particle from decaying to the
standard-model particles.

The mass spectra of KK particles including radiative corrections were
calculated in Ref.~\cite{Cheng:2002iz}. The work~\cite{Cheng:2002iz} 
found that the LKP
was the first KK photon in the broad range on the MUED-model parameter
space, and would be a dark matter candidate. The relic abundance of
the first KK photon dark matter was precisely calculated~\cite{Kakizaki:2005en, 
Burnell:2005hm, Kong:2005hn, Kakizaki:2006dz}. From those analyse, 
the compactification scale
$1/R$ consistent with the cosmological observations is $600
~\text{GeV} \lesssim 1/R \lesssim 1400~\text{GeV}$ and the
standard-model Higgs boson mass is $m_h \lesssim 230$ GeV. From the
indirect low-energy experiments and corresponding analyses, which are
$b \to s \gamma$~\cite{Agashe:2001xt}, $B$ meson
decay~\cite{Buras:2003mk, Haisch:2007vb, Bashiry:2008en}, $B^0$-$\bar
B^0$ oscillation~\cite{Chakraverty:2002qk}, and electroweak precision
measurements~\cite{Gogoladze:2006br}, the lower bound on the
compactification scale is obtained as $1/R \gtrsim 600$ GeV. Thus, the
favorable compactification scale in light of the relic abundance is
free from experimental limits, and despite the simple framework,
the MUED model is phenomenologically successful.

In the following, we assume the first KK photon to be the LKP and
evaluate its elastic scattering cross section with nucleons, which is
relevant to the direct detection of the LKP dark matter. At the
beginning of this section, we explain the parametrization of the MUED
model, and describe the mass spectra of KK particles.  Then, we give
the effective couplings of the LKP with quarks and gluons.  Finally, the
numerical results are shown.

\subsection{Mass spectra  of KK particles}  
\label{subsec:KKmass}

First, we summarize the mass spectra of KK particles relevant to the
direct detection of LKP dark matter.  We employ the minimal
particle contents, {\it i.e.}, gravitons and right-handed neutrinos are
not included in this framework.  Then, the first KK photon
($\gamma^{(1)}$) or the first KK charged Higgs boson ($H^{\pm (1)}$)
becomes the LKP, depending on the compactification scale and 
standard-model Higgs boson mass.  Since the $H^{\pm (1)}$ dark matter
scenario is excluded phenomenologically, we mainly study the parameter
region where $\gamma^{(1)}$ is the LKP unless otherwise mentioned.

The mass of the first KK photon is obtained by diagonalizing the
squared mass matrix for the first KK neutral gauge bosons,
\begin{equation}
\begin{pmatrix}
(1/R)^2 + \delta m_{B^{(1)}}^2 + g_1^2 v^2 /4
& g_1 g_2 v^2 /4
\\[2mm]
g_1 g_2 v^2 /4
& (1/R)^2 + \delta m_{W^{(1)}}^2 + g_2^2 v^2 /4
\end{pmatrix} . 
\label{KKgamma_mat}
\end{equation}
This is written in a ($B^{(1)}$, $W^{3(1)}$) basis, where $B^{(1)}$
and $W^{3(1)}$ are the first KK particles of $U(1)_Y$ and $SU(2)_L$
gauge bosons, respectively.  Here, $g_1$ ($g_2$) represents the $U(1)_Y$
($SU(2)_L$) gauge coupling constant, and $v( \simeq 246~{\rm GeV})$ is
the vacuum expectation value of the standard-model Higgs boson. Radiative
corrections, $\delta m_{B^{(1)}}^2$ and $\delta m_{W^{(1)}}^2$, were
calculated in the work~\cite{Cheng:2002iz} as
\begin{equation}
\begin{split}
   \delta m^2_{B^{(1)}} 
   = - \frac{39}{2} \frac{g_1^2 \zeta(3)}{16 \pi^4 R^2} 
   - \frac{1}{6}  
   \frac{g_1^2}{16 \pi^2 R^2} \ln (\Lambda^2 R^2) , 
\end{split}     \label{cor_B}
\end{equation}
\begin{equation}
\begin{split}
   \delta m^2_{W^{(1)}} 
   = - \frac{5}{2} \frac{g_2^2 \zeta(3)}{16 \pi^4 R^2} 
   + \frac{15}{2} 
   \frac{g_2^2}{16 \pi^2 R^2} \ln (\Lambda^2 R^2) . 
\end{split}     \label{cor_W}
\end{equation}
Here, $\Lambda$ is the ultraviolet cutoff scale of the UED model, which is
ordinarily taken to be $\Lambda \sim
\mathcal{O}(10/R)$~\cite{Cheng:2002iz}.  ($\zeta(n)$ is Riemann's
zeta function.) Since the off-diagonal component is so small in the
mass matrix, the KK photon $\gamma^{(1)}$ is approximately identical
to the weak eigenstate $B^{(1)}$.  Therefore, we consider $B^{(1)}$ as
the $\gamma^{(1)}$, and take the mass of the LKP equal to the mass of
$B^{(1)}$.  On the other hand, the mass of the first KK charged Higgs
boson is given by
\begin{equation}
\begin{split}
   m_{H^{\pm (1)}}^2 
   = (1/R )^2 + m_W^2 + \delta m_{H^{\pm (1)}}^2, 
\end{split}     \label{m_chargedH1}
\end{equation}
\begin{equation}
\begin{split}
   \delta m_{H^{\pm (1)}}^2 
   = \biggl( 
   \frac{3}{4} g_1^2 + \frac{3}{2} g_2^2 - \lambda_h
   \biggr) 
   \frac{1}{16 \pi^2 R^2} 
   \ln (\Lambda^2 R^2) .
\end{split}     \label{m_chargedH}
\end{equation}
Here, $\lambda_h$ is the self-coupling constant of the standard-model Higgs
boson, which is defined by $\lambda_h \equiv m_h^2/v^2$.  When the
standard-model Higgs boson is sufficiently heavy, the radiative correction
$\delta m_{H^{\pm (1)}}^2$ gains a negative contribution so that the
first KK charged Higgs boson is the LKP.

Next, we describe the mass spectra of KK quarks. They are important
since most of the effective couplings of LKP dark matter with
nucleons are very sensitive to the mass-square difference between the
KK quarks and dark matter, which we will see later.  The mass matrix
of the $n$-th KK quarks is written in the weak eigenstate
basis. The up-type quark mass matrix, for instance, is given by
\begin{eqnarray}
-
    \begin{pmatrix}
    \bar{u}^{'(n)} &  \bar{U}^{'(n)} 
    \end{pmatrix}
    \begin{pmatrix}
    -m_{u1}^{(n)} & m_u \\ m_u & m_{u2}^{(n)}
    \end{pmatrix}
    \begin{pmatrix}
    u^{'(n)}  \\ U^{'(n)}
    \end{pmatrix}~,  \label{massm}
\end{eqnarray}
where $u^{'(n)}$ and $U^{'(n)}$ describe the $n$-th KK modes
associated with the $SU(2)_L$ singlet up-type quark and the up-type 
quark in the $SU(2)_L$ doublet, respectively, and $m_u$ is the mass of the
(zero-mode, \textit{i.e.}, standard-model) up-type quark. The down-type
quark mass  is written similarly.  The parameters $m_{u1}^{(n)}$ and
$m_{u2}^{(n)}$ are given as $ m_{u1}^{(n)}=n/R + \delta m_{u^{(n)}}$
and $m_{u2}^{(n)}=n/R + \delta m_{U^{(n)}}$ with the radiative
corrections:
\begin{eqnarray}
   \delta m_{u^{(n)}} 
   &=& \biggl(
   3 g_s^2 + g_1^2 
   \biggr) 
   \frac{n}{R} 
   \frac{\ln (\Lambda^2 R^2)}{16 \pi^2}, 
\nonumber \\
   \delta m_{U^{(n)}} 
   &=& \biggl(
   3 g_s^2 + \frac{27}{16} g_2^2 + \frac{1}{16} g_1^2 
   \biggr) 
   \frac{n}{R} 
   \frac{\ln (\Lambda^2 R^2)}{16 \pi^2}.
\end{eqnarray}
On the other hand, the radiative corrections in the mass matrix of the
$n$-th KK down-type quarks ($d^{\prime (n)}$ and $D^{\prime (n)}$ for
$SU(2)_L$ singlet and in doublet, respectively) are
\begin{eqnarray}
   \delta m_{d^{(n)}} 
   &=& \biggl(
   3 g_s^2 + \frac{1}{4} g_1^2 
   \biggr) 
   \frac{n}{R} 
   \frac{\ln (\Lambda^2 R^2)}{16 \pi^2}, 
\nonumber \\
 \delta m_{D^{(n)}} 
   &=&  \delta m_{U^{(n)}}.
    \label{cor_D}
\end{eqnarray}
The equality in the latter comes from the fact that $D^{\prime (n)}$ and
$U^{\prime (n)}$ make a $SU(2)_L$ doublet.  In addition to these
radiative corrections from gauge couplings, the third-generation KK
quarks also receive the corrections from top Yukawa coupling
$y_t$. They are given by
\begin{eqnarray}
   \delta_{y_t} m_{t^{(n)}} 
   &=& \biggl(
   - \frac{3}{2} y_t^2  
   \biggr) 
   \frac{n}{R} 
   \frac{\ln (\Lambda^2 R^2)}{16 \pi^2} ,
\nonumber \\
   \delta_{y_t} m_{T^{(n)}} 
   &=& \delta_{y_t} m_{B^{(n)}} = 
   \biggl(
   - \frac{3}{4} y_t^2  
   \biggr) 
   \frac{n}{R} 
   \frac{\ln (\Lambda^2 R^2)}{16 \pi^2} ,
\end{eqnarray}
for the $SU(2)$ singlet and doublet, respectively.  The mass eigenstates
of the $n$-th KK quark modes are expressed in linear combinations of
their weak eigenstates. For example, in the up-type quark sector, the
relationship between the mass eigenstates (denoted as $u^{(n)}$ and
$U^{(n)}$) and the weak eigenstates is expressed as follows:
\begin{equation}
    \begin{pmatrix}
    u^{'(n)} \\ U^{'(n)}
    \end{pmatrix}
=
    \begin{pmatrix}
    -\gamma_5 \cos \alpha^{(n)} & \sin \alpha^{(n)}  \\
    \gamma_5 \sin \alpha^{(n)} & \cos \alpha^{(n)}
    \end{pmatrix}
    \begin{pmatrix}
     u^{(n)} \\ U^{(n)}
    \end{pmatrix} .  \label{eigens}
\end{equation}
Then, one obtains the mass eigenvalues from Eqs.~(\ref{eigens}) and
(\ref{massm}) as
\begin{equation}
m_{u^{(n)}/U^{(n)}}=
\sqrt{\biggl(\frac{m_{u1}^{(n)}+m_{u2}^{(n)}}{2}\biggr)^2+m_u^2}
\pm 
\frac{m_{u1}^{(n)}-m_{u2}^{(n)}}{2}
\end{equation}
with the mixing angle
\begin{equation}
\sin 2\alpha^{(n)}_u=\frac{2m_u}{m_u^{(n)}+m_U^{(n)}}.
\end{equation}
The off-diagonal component in Eq.~(\ref{massm}) is small so that the
mass eigenstates are approximately given by the weak ones and the
mixing angle is tiny.  However, the mixing should be kept in finite
when we calculate the scalar-type coupling constant $f_q$ in
Eq.~(\ref{eff_lagq}), since such a type of coupling flips the chirality of
the quark and the mixing is expected to contribute to the coupling.

\begin{figure}[t]
\begin{center}
\parbox{\halftext}{
\includegraphics[width=70mm]{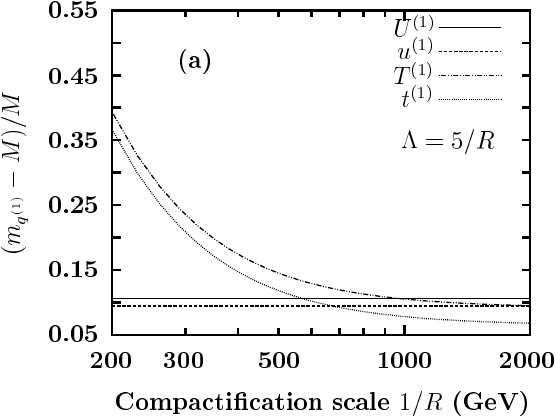}}
\hfill
\parbox{\halftext}{
\includegraphics[width=70mm]{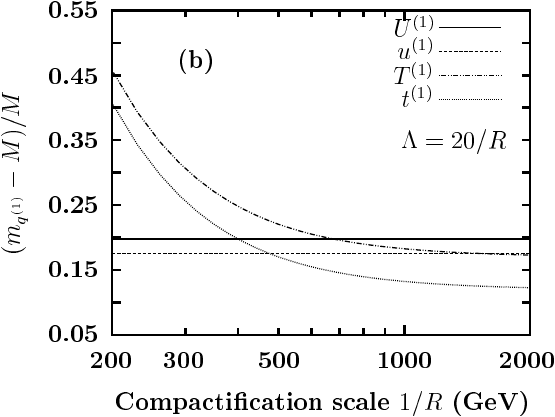}}
\quad \\
\includegraphics[width=70mm]{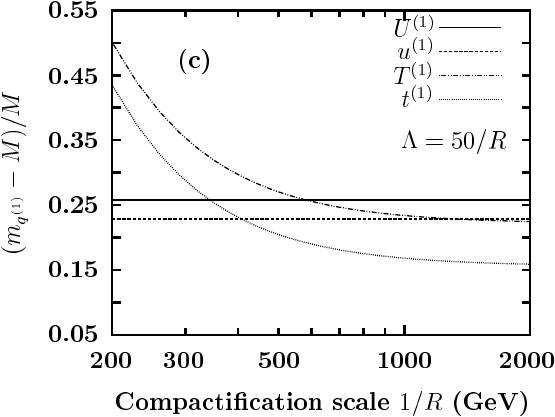}
\caption{Mass degeneracy between LKP dark matter and first KK
  quarks, $(m_{q^{(1)}} - M)/M$, for $\Lambda = 5/R$, $20/R$, and
  $50/R$.  }
\label{Dege}
\end{center}
\end{figure}

Figure~\ref{Dege} shows the mass degeneracy between LKP dark matter
and first KK quarks, $(m_{q^{(1)}} - M)/M$, for $q = U$, $u$, $T$, and
$t$.  We set $\Lambda = 5/R$, $20/R$, and $50/R$ for panels (a), (b),
and (c), respectively.  In this calculation, the gauge coupling
constants are set to be the values at the electroweak scale.  For a 
larger cutoff scale, the mass degeneracy is relaxed owing to the
logarithmic factor in radiative corrections.

\subsection{Scattering with light quarks and gluons}
\begin{figure}
 \begin{center}
   \includegraphics[height=3.5cm]{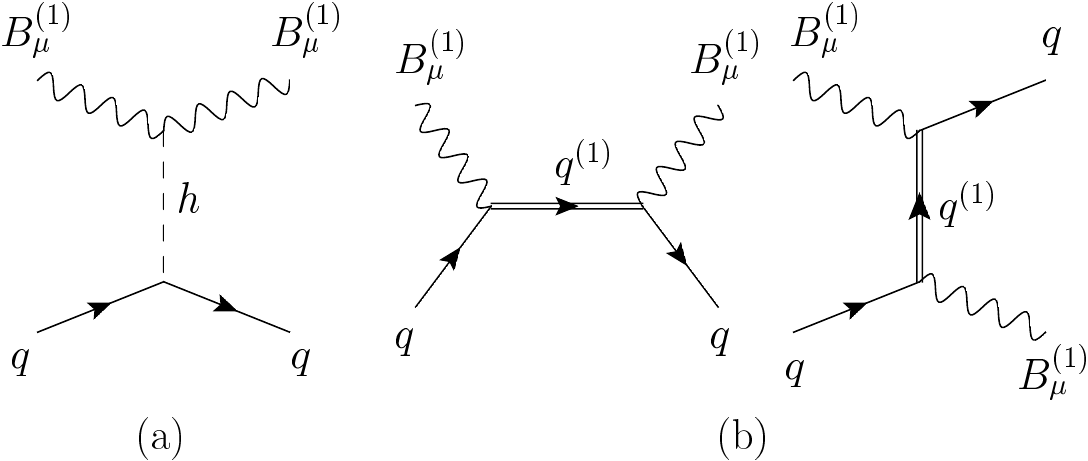}
   \caption{Tree level diagrams for elastic scattering of
     $B^{(1)}$ with light quarks: (a) Higgs boson exchange
     contribution, and (b) KK quark exchange contributions.}
\label{fig:tree}
 \end{center}
\end{figure}

Now, we calculate the scattering amplitude of the LKP with light
quarks.  This process is given at the tree level. Although several authors
have already evaluated the
contribution~\cite{Cheng:2002ej,Servant:2002hb,Arrenberg:2008wy,
  Oikonomou:2006mh}, they mixed the contribution of twist-2-type operators
with the scalar-type one and did not evaluate them correctly.  The
relevant Lagrangian is obtained by taking
$(\psi_1,\psi_2)=(q,Q^{(1)})$ and $(q,q^{(1)})$ in Eq.\
(\ref{eq:simpleL}). Here, $q^{(n)}$ and $Q^{(n)}$ describes the mass
eigenstate of the $n$-th KK quarks, which are the $SU(2)_L$ singlet and
doublet, respectively, and the coupling coefficients are given as (see
Appendix) 
\begin{eqnarray}
a_{Q^{(1)}q} 
&=&-\frac{g_1}{2}(\cos\alpha^{(1)}Y_{\mathrm{qL}}+\sin\alpha^{(1)}Y_{\mathrm{qR}}),
\notag \\
b_{Q^{(1)}q} 
&=&-\frac{g_1}{2}(-\cos\alpha^{(1)}Y_{\mathrm{qL}}+\sin\alpha^{(1)}Y_{\mathrm{qR}}),
\notag \\
a_{q^{(1)}q} 
&=&\frac{g_1}{2}(\sin\alpha^{(1)}Y_{\mathrm{qL}}+\cos\alpha^{(1)}Y_{\mathrm{qR}}),
\notag \\
b_{q^{(1)}q}
&=&\frac{g_1}{2}(-\sin\alpha^{(1)}Y_{\mathrm{qL}}+\cos\alpha^{(1)}Y_{\mathrm{qR}}). 
\label{parameters}
\end{eqnarray}
In the above expression, $Y_{\rm qL}$ and $Y_{\rm qR}$ are
hypercharges of left-handed and right-handed quarks, respectively.
(For instance, $Y_{\rm uL}=\frac{1}{6},Y_{\rm dR}=-\frac{1}{3}$.)  The
mixing angle $\alpha^{(1)}$ is defined in Eq.~(\ref{eigens}).  Those
interaction terms generate diagrams of the first KK quark exchange
(Fig.~\ref{fig:tree}(b)). In addition, there also exists a tree level
scattering process by the standard-model Higgs boson exchange
(Fig.~\ref{fig:tree}(a)).  In the Higgs boson exchange process, not
only the standard-model Higgs boson but $n$-th excited KK Higgs bosons
($n\ge 2$) propagate. However, this effect is so small compared with
that of the standard-model Higgs boson that we ignore the
contribution.

Using the results in the previous section, we derive the coefficients
of the effective Lagrangian in Eq.~(\ref{eff_lagq}) as
\begin{eqnarray}
f_q&=&-\frac{g_1^2}{4m_h^2}-\frac{g_1^2}{4}\biggl[Y^2_{\mathrm{qL}}
\frac{m^2_{Q^{(1)}}}{(m^2_{Q^{(1)}}-M^2)^2}+
Y^2_{\mathrm{qR}}\frac{m^2_{q^{(1)}}}{(m^2_{q^{(1)}}-M^2)^2}\biggr] 
 \nonumber \\ &&
+\frac{g_1^2Y_{\mathrm{qL}}Y_{\mathrm{qR}}}{m_{Q^{(1)}}+m_{q^{(1)}}}
\biggl[\frac{m_{Q^{(1)}}}{m^2_{Q^{(1)}}-M^2}
+\frac{m_{q^{(1)}}}{m^2_{q^{(1)}}-M^2}\biggr], 
\label{fq} \\
d_q&=&\frac{ig^2_1M}{2}\biggl[\frac{Y^2_{\mathrm{qL}}}{m^2_{Q^{(1)}}-M^2}
+\frac{Y^2_{\mathrm{qR}}}{m^2_{q^{(1)}}-M^2}\biggr], \\
g_q&=&-g_1^2M^2\biggl[\frac{Y^2_{\mathrm{qL}}}{(m^2_{Q^{(1)}}-M^2)^2}+
\frac{Y^2_{\mathrm{qR}}}{(m^2_{q^{(1)}}-M^2)^2}\biggr].
\end{eqnarray}
The first term in Eq.~(\ref{fq}) corresponds to the standard-model
Higgs exchange contribution, while the other terms come from the KK
quark exchange processes.

Next,  let us  discuss  the scattering  of  the LKP  with gluons.  The
contribution  of  this scattering  process  has  not  been taken  into
account or  has not been properly  calculated before. As  we will see,
however, it is  not negligible. This scattering process  is induced by
one-loop  level.  The  diagrams that contribute to the  scattering with
gluons are  categorized into three types:  (i) quark and  the first KK
quark  contribution,  (ii) higher  KK  quark  contribution, and  (iii)
standard-model Higgs boson exchange contribution, such that
\begin{eqnarray}
f_G=f_G^{\rm (i)}+f_G^{\rm (ii)}+f_G^{\rm (iii)}.
\end{eqnarray}

\begin{figure}[t]
 \begin{center}
   \includegraphics[height=4cm]{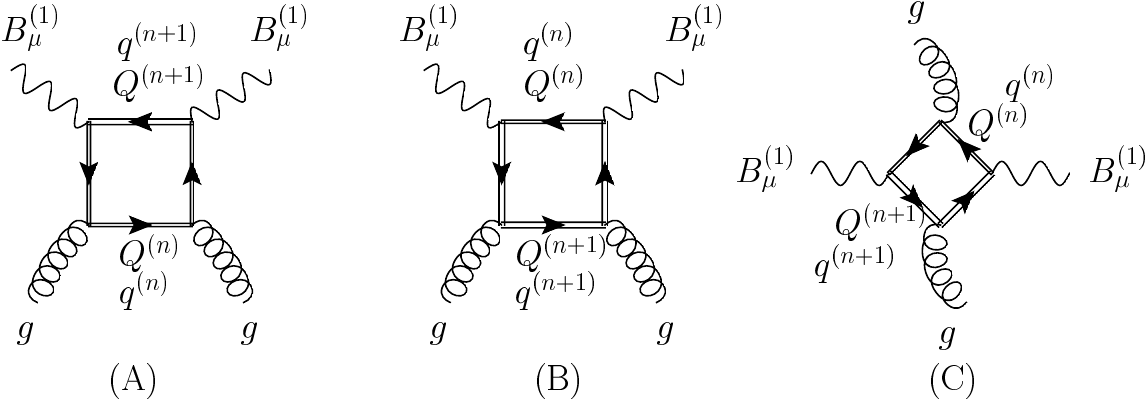}
   \caption{One-loop diagrams for the effective interaction of
     $B^{(1)}$ with gluons via KK quarks. Diagram (A) gives a 
     long-distance contribution, whereas (B) and (C) 
     give short-distance ones.}
\label{fig:qloop}
 \end{center}
\end{figure}
Let us first closely look at contributions (i) and (ii). They are
induced from the diagrams depicted in Fig.~\ref{fig:qloop}. The
type (i) contribution is given by the interaction Lagrangian, which was
introduced in the previous tree level calculation ({\it i.e.},
$(\psi_1,\psi_2)=(q,q^{(1)})$ and $(q,Q^{(1)})$).  The relevant
Lagrangian for the type (ii) contribution, on the other hand, is given by
taking $(\psi_1,\psi_2)=(q^{(n)},q^{(n+1)})$, $(Q^{(n)},Q^{(n+1)})$,
$(q^{(n)},Q^{(n+1)})$, and $(Q^{(n)},q^{(n+1)})$ $(n\ge 1)$ in
Eq.~(\ref{eq:simpleL}), with the couplings,
\begin{eqnarray}
 a_{q^{(n+1)}q^{(n)}}=-\frac{g_1}{\sqrt{2}}
[Y_{\mathrm{qL}}\sin\alpha^{(n)}\sin\alpha^{(n+1)}
+Y_{\mathrm{qR}}\cos\alpha^{(n)}\cos\alpha^{(n+1)}], \nonumber \\
 a_{Q^{(n+1)}Q^{(n)}}=-\frac{g_1}{\sqrt{2}}
[Y_{\mathrm{qL}}\cos\alpha^{(n)}\cos\alpha^{(n+1)}
+Y_{\mathrm{qR}}\sin\alpha^{(n)}\sin\alpha^{(n+1)}], \nonumber \\
 b_{Q^{(n+1)}q^{(n)}}=-\frac{g_1}{\sqrt{2}}
[Y_{\mathrm{qL}}\sin\alpha^{(n)}\cos\alpha^{(n+1)}
-Y_{\mathrm{qR}}\cos\alpha^{(n)}\sin\alpha^{(n+1)}], \nonumber \\
 b_{q^{(n+1)}Q^{(n)}}=-\frac{g_1}{\sqrt{2}}
[Y_{\mathrm{qL}}\cos\alpha^{(n)}\sin\alpha^{(n+1)}
-Y_{\mathrm{qR}}\sin\alpha^{(n)}\cos\alpha^{(n+1)}], \nonumber \\
b_{q^{(n+1)}q^{(n)}}= b_{Q^{(n+1)}Q^{(n)}}=
 a_{Q^{(n+1)}q^{(n)}}= a_{q^{(n+1)}Q^{(n)}}=0.
\end{eqnarray}
(For the derivation, see Appendix.)

As we discussed in \S\ref{sec:EffLag}, there exists the
long-distance contribution when the quark runs in the loop. This is the
case for the type (i) contribution.  Thus, in the calculation of the type (i)
contribution, the diagrams in which light quarks run in the loop should
not be included.  Then, the contribution to the scalar effective
coupling to the gluon is written as
\begin{eqnarray}
 f_G^{\rm (i)}=
\frac{\alpha_s}{4\pi}\sum_{q=c,b,t} c_q
\biggl[
 (a^2_{Q^{(1)}q}+b^2_{Q^{(1)}q})f^{\rm (A)}_+(M;m_q,m_{Q^{(1)}})
+(a^2_{Q^{(1)}q}-b^2_{Q^{(1)}q})f^{\rm (A)}_-(M;m_q,m_{Q^{(1)}})\nonumber \\ 
+(a^2_{q^{(1)}q}+b^2_{q^{(1)}q})f^{\rm (A)}_+(M;m_q,m_{q^{(1)}})
+(a^2_{q^{(1)}q}-b^2_{q^{(1)}q})f^{\rm (A)}_-(M;m_q,m_{q^{(1)}})
 \biggr] \nonumber \\
+\frac{\alpha_s}{4\pi}\sum_{q={\rm all}} \sum_{{\rm I}={\rm B,C}}
\biggl[
 (a^2_{Q^{(1)}q}+b^2_{Q^{(1)}q})f^{\rm(I)}_+(M;m_q,m_{Q^{(1)}})
+(a^2_{Q^{(1)}q}-b^2_{Q^{(1)}q})f^{\rm (I)}_-(M;m_q,m_{Q^{(1)}}) \nonumber \\ 
+(a^2_{q^{(1)}q}+b^2_{q^{(1)}q})f^{\rm (I)}_+(M;m_q,m_{q^{(1)}})
+(a^2_{q^{(1)}q}-b^2_{q^{(1)}q})f^{\rm (I)}_-(M;m_q,m_{q^{(1)}})
 \biggr].
\nonumber\\
\label{quarkl}
\end{eqnarray}
On the other hand, for the type (ii) contribution, all the diagrams are 
short-distance contributions.  Then,
\begin{eqnarray}
f_G^{\rm (ii)}
=
\frac{\alpha_s}{4\pi}\sum_{q={\rm all}}\sum_{{\rm I}={\rm A,B,C}}
\biggl\{
a_{q^{(n+1)}q^{(n)}}^2\bigl[f^{\rm (I)}_+(M;m_{q^{(n)}},m_{q^{(n+1)}})
+f^{\rm (I)}_-(M;m_{q^{(n)}},m_{q^{(n+1)}})\bigr] \nonumber \\
+a_{Q^{(n+1)}Q^{(n)}}^2\bigl[f^{\rm (I)}_+(M;m_{Q^{(n)}},m_{Q^{(n+1)}})
+f^{\rm (I)}_-(M;m_{Q^{(n)}},m_{Q^{(n+1)}})\bigr] \nonumber \\
+b_{Q^{(n+1)}q^{(n)}}^2\bigl[f^{\rm (I)}_+(M;m_{q^{(n)}},m_{Q^{(n+1)}})
-f^{\rm (I)}_-(M;m_{q^{(n)}},m_{Q^{(n+1)}})\bigr] \nonumber \\
+b_{q^{(n+1)}Q^{(n)}}^2\bigl[f^{\rm (I)}_+(M;m_{Q^{(n)}},m_{q^{(n+1)}})
-f^{\rm (I)}_-(M;m_{Q^{(n)}},m_{q^{(n+1)}})\bigr]
\biggr\}.\nonumber \\
\label{kkquarkl}
\end{eqnarray}

\begin{figure}[t]
 \begin{center}
   \includegraphics[height=4cm]{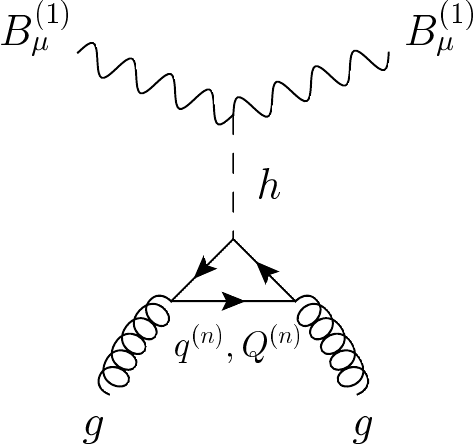}
   \caption{One-loop diagram for the effective interaction of
     $B^{(1)}$ with gluons through the exchange of standard-model Higgs
     boson.}
\label{fig:hloop}
 \end{center}
\end{figure}

Finally, we discuss the type (iii) contribution. The standard-model Higgs
boson is coupled with the standard-model quarks and also higher KK
quarks, and then the triangle loop diagrams of heavy quarks and KK
quarks contribute to the effective scalar coupling to gluons.  The
corresponding diagram is shown in Fig.~\ref{fig:hloop}.  In the
triangle loop, all the higher KK quarks run in the loop, in addition to
heavy quarks.  Then, we obtain the standard-model Higgs exchange
contribution as
\begin{eqnarray}
f_G^{\rm (iii)}&=&\frac{g^2_1\alpha_s}{48\pi
 m^2_h}\Bigl[(c_c+c_b+c_t) +m_t c_t
 \sum_n \sin 2\alpha^{(n)}\Bigl(\frac{1}{m_{t^{(n)}}}+\frac{1}{m_{T^{(n)}}} 
 \Bigr) \Bigr]  \notag \\ 
&=&\frac{g^2_1\alpha_s}{48\pi m^2_h}
\Bigl[(c_c+c_b+c_t)+c_t\sum_n \frac{2m_t^2}{m_{t^{(n)}}m_{T^{(n)}}}
\Bigr].
\label{higgsl}
\end{eqnarray} 
In this calculation, we ignore quark masses except that of the top quark.

\subsection{Results}

\begin{figure}[t]
 \begin{center}
   \includegraphics[scale=0.75]{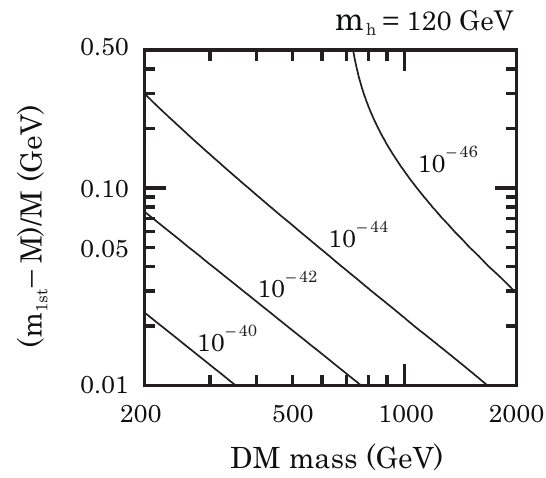}
   \includegraphics[scale=0.75]{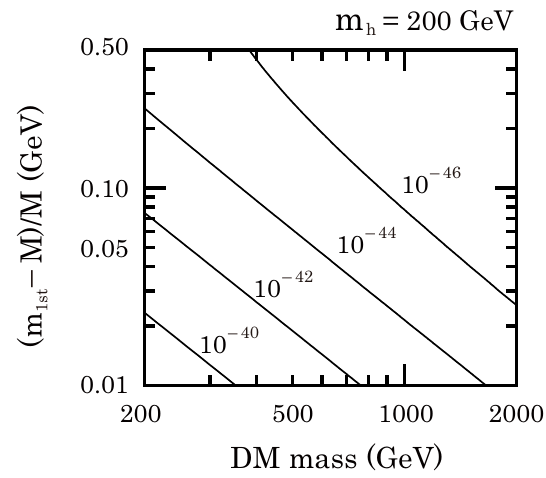}
   \quad
   \\
   \includegraphics[scale=0.75]{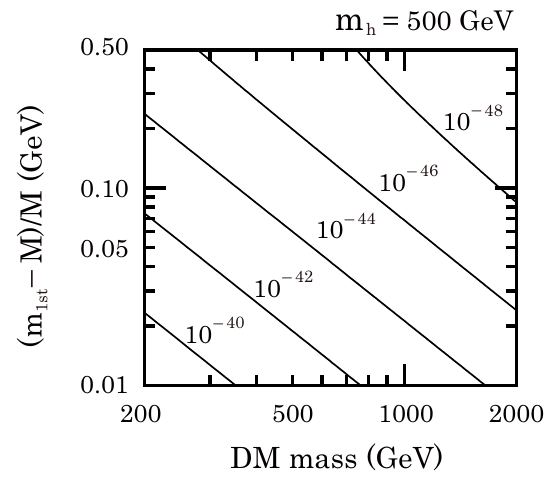}
   \caption{Spin-independent cross section with a proton
     on the plane, LKP mass vs.~mass degeneracy $(m_{\rm 1st}-M)/M$. 
     Here, we take $m_h=120~{\rm GeV}$ (upper left), 200~GeV (upper right), 
     and 500~GeV (bottom). Lines correspond to the contours 
     $\sigma_{\rm SI}=10^{-40}~{\rm cm}^2$, $10^{-42}~{\rm cm}^2$,
     $10^{-44}~{\rm cm}^2$, $10^{-46}~{\rm cm}^2$, and $10^{-48}~{\rm
       cm}^2$ from bottom to top.}
\label{fig:SIsigmaMh120}
 \end{center}
\end{figure}

Now, we are at the point of showing the numerical results of the cross
section.  At the beginning, let us examine the case in which $n$-th KK
modes with $n\ge 2$ are neglected so that we see features of the cross
section.  Also, it is worthwhile when one applies the formulae for
other vector dark matter scenarios. In Fig.~\ref{fig:SIsigmaMh120},
the spin-independent cross section of the LKP with a proton
 ($\sigma_{\rm SI}$) is depicted.  Here, we set 
$m_{q^{(1)}}=m_{Q^{(1)}}\equiv m_{\rm 1st}$ for simplicity, except for
the masses of the first KK top quarks, for which we take
$m^2_{t^{(1)}}=m^2_{T^{(1)}}=m^2_{\rm 1st}+m^2_t$ ($m_t$ is top quark
mass), and show the contours of the cross section on the plane, the
LKP mass vs the mass degeneracy between the first excited KK quarks
and the LKP, $(m_{\rm 1st}-M)/M$.  The standard-model Higgs boson mass
is set as $m_h=120~{\rm GeV}$, 200 GeV, and 500 GeV.  From the figure,
it is found that the cross section is enhanced when the first KK quark
and LKP masses are degenerate, as is expected. Such a behavior was
already discussed in the previous works.  However, with the complete
calculation that has been carried out here, we have discovered that the
spin-independent cross section is larger than the former theoretical
predictions. In the case of $m_h=120~{\rm GeV}$ and the mass
degeneracy of 10\%, for instance, the cross section ranges from $3.5
\times 10^{-43}~{\rm cm}^2$ to $1.3 \times 10^{-46}~{\rm cm}^2$ when
the LKP mass is from $200~{\rm GeV}$ to 1 TeV.  This result is larger
than those in the previous works
\cite{Cheng:2002ej,Servant:2002hb,Arrenberg:2008wy} up to about
a factor of ten. In our calculation, we have found that the
twist-2-type operator contribution, which had not been taken into
account correctly in the previous works, dominates the
effective coupling when $M \lesssim O({\rm TeV})$ and enhances the
cross section.  In the parameter region $M \gtrsim O({\rm TeV})$, on
the other hand, the standard-model Higgs boson contribution (from both
tree-  and one-loop-level diagrams) dominates the effective
coupling. Then, the cross section is determined using the standard-model
Higgs boson mass in the parameter region.

\begin{figure}[t]
 \begin{center}
   \includegraphics[scale=0.5]{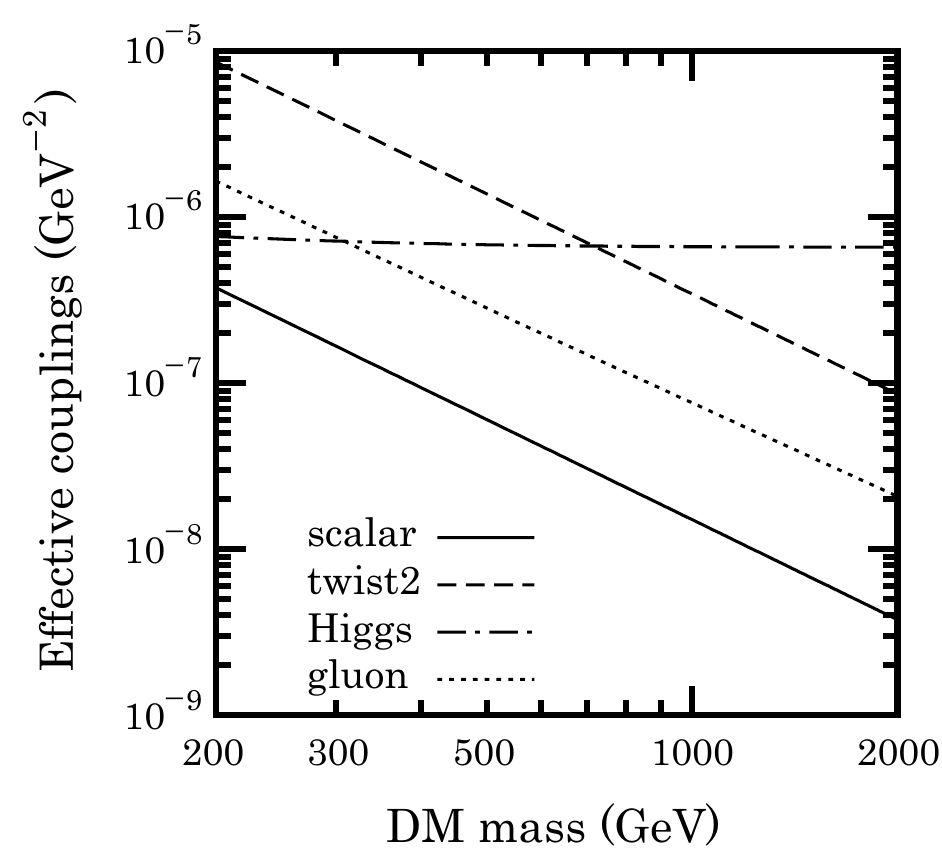}
   \caption{Each contribution in effective coupling $f_N/m_N$ given in
     Eq.~(\ref{f}). Here, we set $m_h=120~{\rm GeV}$ and $(m_{\rm
       1st}-M)/M=0.1$.  ``scalar'',``twist-2'', and ``gluon''
     correspond to the first, second, and third terms in Eq.~(\ref{f})
     (except for standard-model Higgs exchange contribution),
     respectively, and standard-model Higgs exchange contribution,
     including tree and one-loop levels, is denoted as ``Higgs''. }
\label{fig:fN_10}
 \end{center}
\end{figure}
\begin{figure}[t]
\begin{center}
  \includegraphics[scale=0.6]{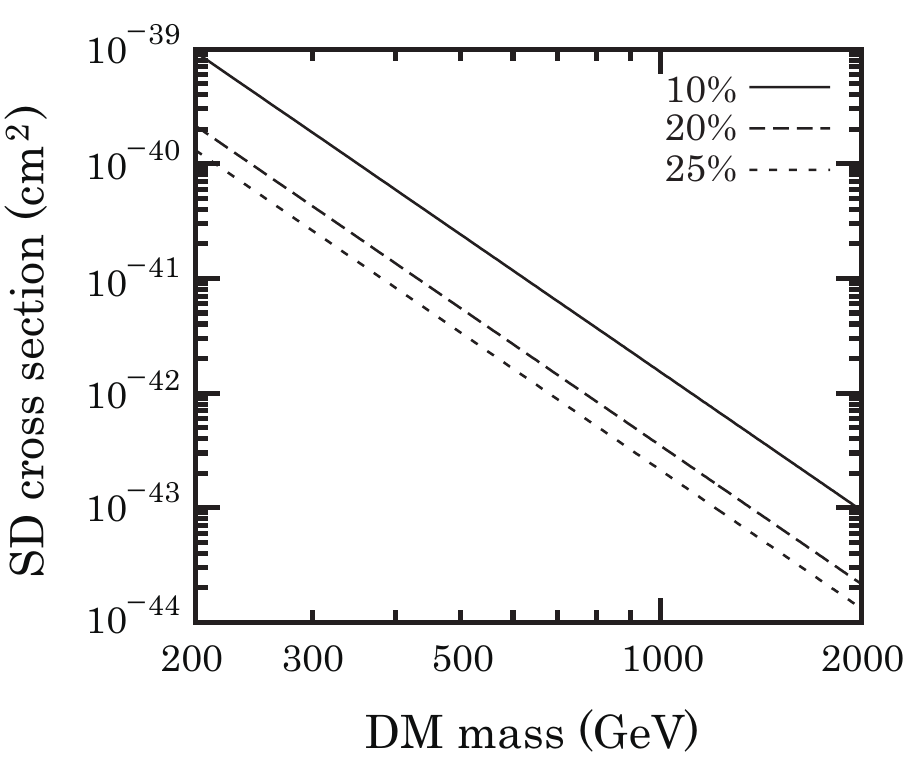}
  \caption{Spin-dependent cross section with a proton.  
    Here, we set $(m_{\rm 1st}-M)/M=0.1$, 0.2, and 0.25 from top to bottom. }
\label{spin-dependent}
\end{center}
\end{figure}

To clarify the behavior of the cross section, we also plot
each contribution in the effective coupling $f_N$ in
Fig.~\ref{fig:fN_10}.  Here, we set $m_h=120~{\rm GeV}$ and 10\% of the
mass degeneracy.  In the figure, ``scalar'',``twist-2'', and ``gluon''
correspond to the first, second, and third terms in Eq.~(\ref{f})
(except for the standard-model Higgs exchange contribution),
respectively, and the standard-model Higgs exchange contribution
(including tree and one-loop levels) is denoted as``Higgs''.  We have
found that all the contributions have the same (negative) sign so that
they are constructive. As a consequence, the cross section is enhanced
in a wide parameter region. In the figure, it is seen that the
twist-2 contribution dominates the effective coupling when $M \lesssim
O({\rm TeV})$. The gluon contribution is subleading, but not
negligible at all.  On the other hand, the standard-model Higgs boson
contribution dominates the effective coupling when $M \gtrsim
O({\rm TeV})$.

For completeness, we show the spin-dependent scattering cross section. 
The spin-dependent interaction is generated by KK quark exchange at the 
tree level, and it is found that the cross section is enhanced when the
first KK quark and LKP masses are degenerate as expected from
Eq.~(\ref{tree_general_2}). In Fig.~\ref{spin-dependent}, the
spin-dependent scattering cross section with a proton 
is shown as a function of the LKP mass assuming the degeneracies are 10\%, 
20\%, and 25\%. We found that the spin-dependent 
scattering cross section is consistent with 
Refs.~\cite{Cheng:2002ej,Servant:2002hb,Arrenberg:2008wy}.

\begin{figure}[t]
  \begin{center}
    \includegraphics[scale=0.6]{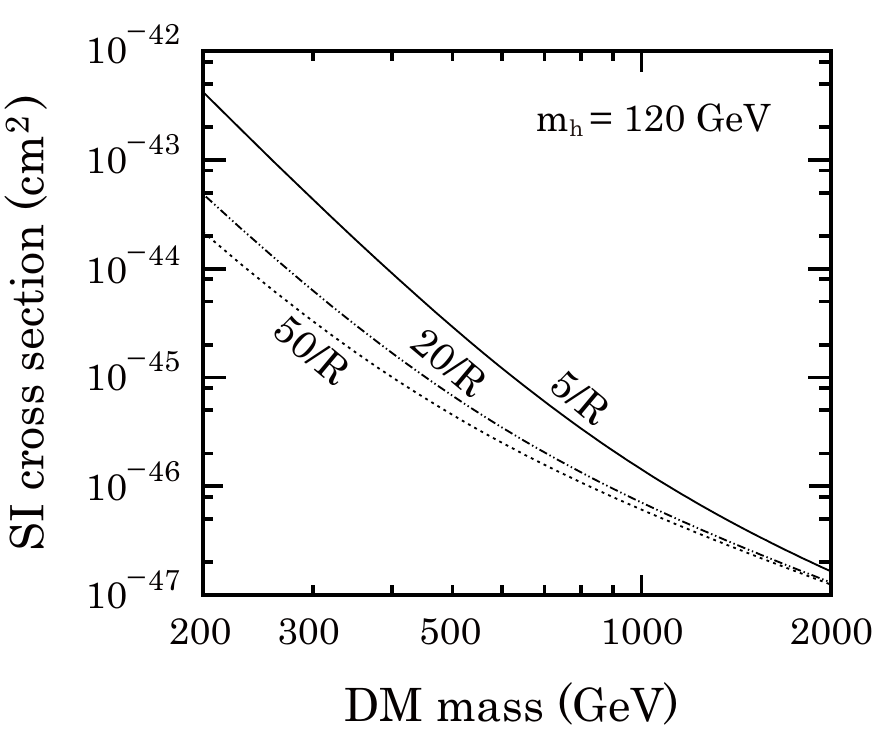}
  \end{center}
  \caption{Spin-independent cross section with a proton 
    for $m_h = 120$GeV. Each line corresponds to $\Lambda = 5/R$, $20/R$, 
    and $50/R$, respectively. }
  \label{mhfixed}
\end{figure}

Next, we discuss a more realistic situation for the direct detection of
the LKP dark matter in the MUED model, including higher KK mode
contributions. For the cutoff scale $\Lambda = n/R$, the KK particles
up to the $n$-th mode are included in the calculations.  Furthermore,
we employ the radiative corrections for the spectra of KK particles
discussed in \S\ref{subsec:KKmass}.  To see the importance of the 
higher KK mode contribution, we show the cutoff dependence of the
spin-independent cross section with a proton by 
taking $\Lambda = 5/R$, $20/R$, and $50/R$ in Fig.~\ref{mhfixed}. 
In the figure, we take $m_h = 120$ GeV.  For a larger cutoff scale, 
although a larger number of KK modes could contribute to the cross 
section, the cross section tends to be smaller than that for the smaller 
cutoff scale. This is because the large cutoff scale makes the mass 
degeneracy relax owing to the radiative corrections (as shown in 
Fig.~\ref{Dege}), and this reductive effect dominates over the increasing 
effect from higher KK mode contributions.

\begin{figure}[t]
\begin{center}
\includegraphics[scale=0.6]{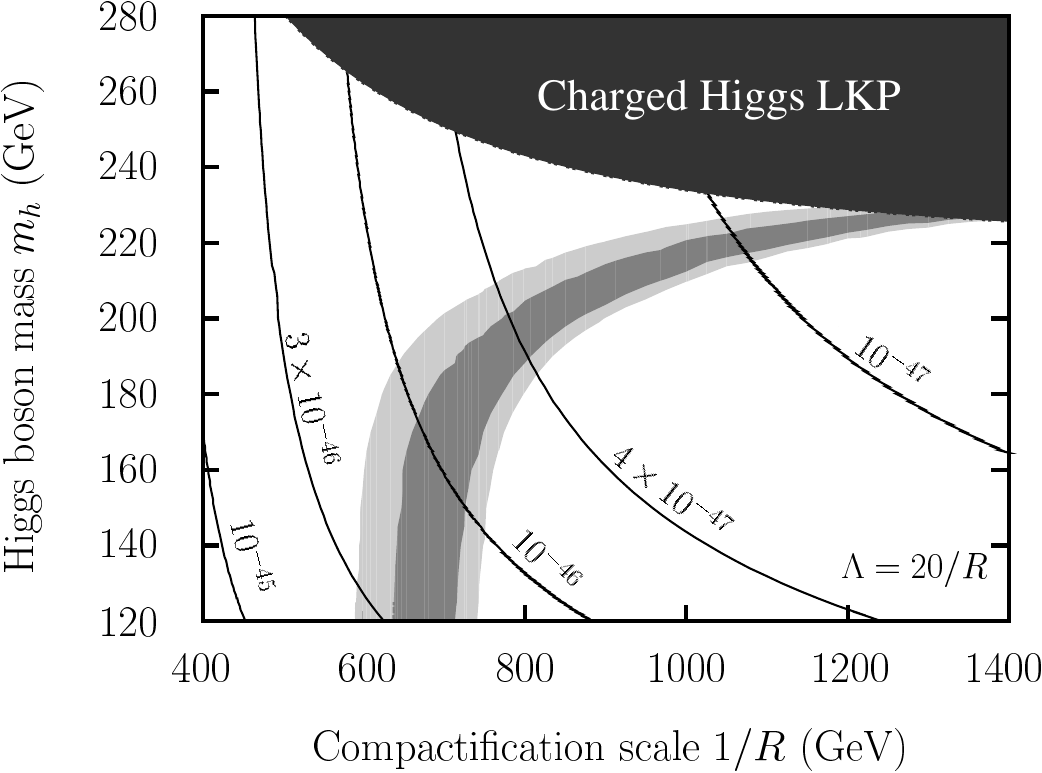}
\caption{Contour plot of relic abundance consistent with WMAP result
  and spin-independent cross section in the ($1/R$, $m_h$) plane for
  $\Lambda = 20/R$. The dark (light) gray region shows the $1\sigma$
  ($2\sigma$) allowed region in light of relic abundance. Each line
  corresponds to a spin-independent scattering cross section with 
  a proton $\sigma_\text{SI} = 10^{-45}~{\rm cm}^2$, 
  $3\times 10^{-46}~{\rm cm}^2$, $10^{-46}~{\rm cm}^2$, $4 \times 
  10^{-47}~{\rm cm}^2$, and $10^{-47}~{\rm cm}^2$. }
\label{abu_DDrate}
\end{center}
\end{figure}

Finally, we discuss the feasibility of the direct detection of KK
photon dark matter for the parameter space in which its relic
abundance can be reproduced in accordance with the WMAP result. We
plot the relationship between the spin-independent scattering 
cross section with a proton and the relic 
abundance in Fig.~\ref{abu_DDrate}.  The charged Higgs boson is the 
LKP in the upper right region, and hence this region is excluded.  
The dark (light) gray region shows the $1\sigma$ ($2\sigma$) 
allowed region from the WMAP observational result, which is
the result calculated in Ref~\cite{Kakizaki:2006dz}.  Each line
corresponds to the spin-independent cross section $\sigma_\text{SI} =
10^{-45}~\text{cm}^2$, $3\times 10^{-46}~\text{cm}^2$,
$10^{-46}~\text{cm}^2$, $4\times 10^{-47}~\text{cm}^2$, and
$10^{-47}~\text{cm}^2$.  
The spin-independent cross section ranges from about $3 \times 
10^{-46}~\text{cm}^2$ to $5 \times 10^{-48}~\text{cm}^2$ on the 
allowed region of the relic abundance, and are just below the present 
experimental bounds. The ongoing and
future experiments for the direct detection of dark matter would reach
sensitivities in this range.

\section{Conclusions}
\label{sec:conc}

In this paper, we assume that WIMP dark matter is composed of
vector particles and have evaluated an elastic cross section with
nucleons in the direct detection experiments. The vector dark matter is
predicted in models beyond the standard model, such as the KK photon in
the UED model and the $T$-odd heavy photon in the Littlest Higgs model
with $T$-parity. On the other hand, however, the cross section had not
yet been consistently evaluated, even at the leading order of
$\alpha_s$. We have derived the general formulae for the cross section
of general vector dark matter with nucleons.

As an application of our formulae, we discussed the direct detection
of the first KK photon in the MUED model. We found that the cross
section is larger than those in the previous works by up to a factor
of 10.  We showed that the spin-independent cross section of the KK
photon with a proton ranges from nearly $ 3 \times 
10^{-46}~{\rm cm}^2$ to $5 \times 10^{-48}~{\rm cm}^2$ on the 
parameter region with the thermal relic abundance consistent with the 
cosmological observations. The future direct detection experiments with 
ton-scale detectors might cover the range.
~\\~\\

{\it Note Added:} While preparing the manuscript, we became aware of a
paper by G.~Belanger, M.~Kakizaki and A.~Pukhov~\cite{Belanger:2010yx}. 
In that paper, the elastic cross section of the
KK photon with nucleons in the MUED model was calculated using 
microOMEGA \cite{Belanger:2006is}. In the calculation, the tree level
contribution was included in the scattering process whereas the one-loop
contributions were incomplete.

\section{Acknowledgments}
We thank Alexander Pukhov for pointing out errors in
Eqs.~\eqref{eq:fplblimit} and \eqref{eq:fplclimit} in the previous
versions of this paper. 
The work was supported in part by Grants-in-Aid from the Ministry of
Education, Culture, Sports, Science and Technology (MEXT), Government of
Japan, No. 20244037, No. 2054252, and No. 2244021 (J.H.).  The work of
J.H. is also supported by the World Premier International Research
Center Initiative (WPI Initiative), MEXT, Japan. This work was also
supported in part by the U.S. Department of Energy under contract
No. DE-FG02-92ER40701, and by the Gordon and Betty Moore Foundation
(K.I.).

\appendix

\section{Relevant Feynman rules}
In this Appendix, we collect parts of the Lagrangian in the MUED model. 

\begin{itemize}
\item Standard-model Higgs boson coupling with the first KK
 particle of $U(1)_Y$ gauge boson
\begin{equation}
\mathcal{L}_{BBh^0}=\frac{1}{4}g^{2}_1 v h^0B^{(1)}_{\mu}B^{(1)\mu}
\end{equation}

\item Standard-model Higgs boson coupling with $n$-th KK
  quarks
\begin{eqnarray}
  \mathcal{L}_{q^{(n)}q^{(n)}h^0}=-\frac{m_q}{v}h^0\sum_{n=1}
  \biggl[\sin2\alpha^{(n)}\bigl(\bar{Q}^{(n)}Q^{(n)}+
  \bar{q}^{(n)}q^{(n)}\bigr)
  \nonumber\\
  +\cos2\alpha^{(n)}\bigl(\bar{q}^{(n)}\gamma_5Q^{(n)}
  -\bar{Q}^{(n)}\gamma_5q^{(n)}\bigr)\biggr]
\end{eqnarray}

\item Coupling of the first KK particle of the $U(1)_Y$ gauge boson with
  the first KK quarks and quarks
\begin{eqnarray}
\mathcal{L}_{q^{(n)}qB}
=-g_1\bar{q}\gamma^{\mu}\bigl[\cos\alpha^{(1)}Y_{\mathrm{qL}}P_{\mathrm{L}}
+\sin\alpha^{(1)}Y_{\mathrm{qR}}P_{\mathrm{R}}\bigr]Q^{(1)}B^{(1)}_{\mu} \notag \\
-g_1\bar{Q}^{(1)}\gamma^{\mu}\bigl[\cos\alpha^{(1)}Y_{\mathrm{qL}}P_{\mathrm{L}}
+\sin\alpha^{(1)}Y_{\mathrm{qR}}P_{\mathrm{R}}\bigr]qB^{(1)}_{\mu} \notag \\
+g_1\bar{q}\gamma^{\mu}\bigl[\sin\alpha^{(1)}Y_{\mathrm{qL}}P_{\mathrm{L}}
+\cos\alpha^{(1)}Y_{\mathrm{qR}}P_{\mathrm{R}}\bigr]q^{(1)}B^{(1)}_{\mu} \notag \\
+g_1\bar{q}^{(1)}\gamma^{\mu}\bigl[\sin\alpha^{(1)}Y_{\mathrm{qL}}P_{\mathrm{L}}
+\cos\alpha^{(1)}Y_{\mathrm{qR}}P_{\mathrm{R}}\bigr]q B^{(1)}_{\mu}
\end{eqnarray}

\item Coupling of the first KK particle of the $U(1)_Y$ gauge boson  with the
$n$-th and $(n+1)$-th KK quarks 
\begin{eqnarray}
\mathcal{L}_{q^{(n)}q^{(n+1)}B}=-\frac{g_1}{\sqrt{2}}
B_{\mu}^{(1)}\sum_{n=1}\bigl
[\bigl(Y_{\mathrm{qL}}\sin\alpha^{(n)}\sin\alpha^{(n+1)}
+Y_{\mathrm{qR}}\cos\alpha^{(n)}\cos\alpha^{(n+1)}\bigr) \notag \\
\times\bigl(\bar{q}^{(n+1)}\gamma^{\mu}q^{(n)}
+\bar{q}^{(n)}\gamma^{\mu}q^{(n+1)}\bigr) \notag \\
+\bigl(Y_{\mathrm{qL}}\cos\alpha^{(n)}\cos\alpha^{(n+1)} 
+Y_{\mathrm{qR}}\sin\alpha^{(n)}\sin\alpha^{(n+1)}\bigr) \notag \\
\times\bigl(\bar{Q}^{(n+1)}\gamma^{\mu}Q^{(n)}
+\bar{Q}^{(n)}\gamma^{\mu}Q^{(n+1)}\bigr) \notag \\
+\bigl(Y_{\mathrm{qL}}\sin\alpha^{(n)}\cos\alpha^{(n+1)}
-Y_{\mathrm{qR}}\cos\alpha^{(n)}\sin\alpha^{(n+1)}\bigr) \notag \\
\times\bigl(\bar{Q}^{(n+1)}\gamma^{\mu}\gamma_5q^{(n)}
+\bar{q}^{(n)}\gamma^{\mu}\gamma_5Q^{(n+1)}\bigr) \notag \\
+\bigl(Y_{\mathrm{qL}}\cos\alpha^{(n)}\sin\alpha^{(n+1)}
-Y_{\mathrm{qR}}\sin\alpha^{(n)}\cos\alpha^{(n+1)}\bigr) \notag \\
\times\bigl(\bar{q}^{(n+1)}\gamma^{\mu}\gamma_5Q^{(n)}
+\bar{Q}^{(n)}\gamma^{\mu}\gamma_5q^{(n+1)}\bigr) \bigr]\nonumber \\
\end{eqnarray}

\end{itemize}


\begin{thebibliography}{99}
\bibitem{Hinshaw:2008kr}
  G.~Hinshaw {\it et al.}  [WMAP Collaboration],
  Astrophys.\ J.\ Suppl.\  {\bf 180} (2009), 225.

\bibitem{Servant:2002aq}
  G.~Servant and T.~M.~P.~Tait,
  Nucl.\ Phys.\  B {\bf 650} (2003), 391.

\bibitem{Cheng:2002ej}
  H.~C.~P.~Cheng, J.~L.~Feng and K.~T.~Matchev,
  Phys.\ Rev.\ Lett.\  {\bf 89} (2002), 211301.

\bibitem{Appelquist:2000nn}
  T.~Appelquist, H.~C.~Cheng and B.~A.~Dobrescu,
  Phys.\ Rev.\  D {\bf 64}  (2001), 035002.

\bibitem{Cheng:2002iz}
  H.~C.~Cheng, K.~T.~Matchev and M.~Schmaltz,
  Phys.\ Rev.\  D {\bf 66} (2002), 036005.

\bibitem{Hubisz:2004ft}
   J.~Hubisz and P.~Meade,
   Phys.\ Rev.\  D {\bf 71} (2005), 035016.

\bibitem{Birkedal:2006fz}
  A.~Birkedal, A.~Noble, M.~Perelstein and A.~Spray,
  Phys.\ Rev.\  D {\bf 74} (2006), 035002.

\bibitem{asano}
  M.~Asano, S.~Matsumoto, N.~Okada and Y.~Okada,
  Phys.\ Rev.\  D {\bf 75} (2007), 063506.

\bibitem{ArkaniHamed:2001nc}
  N.~Arkani-Hamed, A.~G.~Cohen and H.~Georgi,
  Phys.\ Lett.\ B {\bf 513} (2001), 232;
  N.~Arkani-Hamed, A.~G.~Cohen, E.~Katz,
  A.~E.~Nelson, T.~Gregoire and J.~G.~Wacker,
  JHEP {\bf 0208} (2002), 021;
   N.~Arkani-Hamed,
  A.~G.~Cohen, E.~Katz and A.~E.~Nelson,
  JHEP {\bf 0207} (2002), 034.

\bibitem{Cheng:2004yc}
  H.~C.~Cheng and I.~Low,
  JHEP {\bf 0408} (2004), 061;
  I.~Low,
  JHEP {\bf 0410} (2004), 067;

\bibitem{Aprile:2011hi}
  E.~Aprile {\it et al.}  [XENON100 Collaboration],
  arXiv:1104.2549 [astro-ph.CO].

\bibitem{Hisano:2010fy}
  J.~Hisano, K.~Ishiwata and N.~Nagata,
  Phys.\ Lett.\  B {\bf 690} (2010), 311.

\bibitem{Hisano:2010ct}
  J.~Hisano, K.~Ishiwata and N.~Nagata,
  Phys.\ Rev.\ D {\bf 82}  (2010), 115007.

\bibitem{Servant:2002hb}
  G.~Servant and T.~M.~P.~Tait,
  New J.\ Phys.\  {\bf 4} (2002), 99.

\bibitem{Arrenberg:2008wy}
  S.~Arrenberg, L.~Baudis, K.~Kong, K.~T.~Matchev and J.~Yoo,
  Phys.\ Rev.\  D {\bf 78} (2008), 056002.

\bibitem{Drees:1993bu}
  M.~Drees and M.~Nojiri,
  Phys.\ Rev.\  D {\bf 48} (1993), 3483.

\bibitem{Politzer:1980me}
  H.~D.~Politzer,
  Nucl.\ Phys.\  B {\bf 172} (1980), 349.

\bibitem{Corsetti:2000yq}
  A.~Corsetti and P.~Nath,
  Phys.\ Rev.\  D {\bf 64} (2001), 125010.
  [arXiv:hep-ph/0003186].

\bibitem{Ohki:2008ff}
  H.~Ohki {\it et al.},
  Phys.\ Rev.\  D {\bf 78} (2008), 054502.

\bibitem{Cheng:1988im}
  H.~Y.~Cheng,
  Phys.\ Lett.\  B {\bf 219}  (1989), 347.

\bibitem{Adams:1995ufa}
  D.~Adams {\it et al.}  [Spin Muon Collaboration],
  Phys.\ Lett.\  B {\bf 357}  (1995),  248.


\bibitem{Pumplin:2002vw}
  J.~Pumplin, D.~R.~Stump, J.~Huston, H.~L.~Lai, P.~M.~Nadolsky and W.~K.~Tung,
  JHEP {\bf 0207} (2002), 012.

\bibitem{Shifman:1978zn}
  M.~A.~Shifman, A.~I.~Vainshtein and V.~I.~Zakharov,
  Phys.\ Lett.\  B {\bf 78} (1978), 443.

\bibitem{Djouadi:2000ck}
  A.~Djouadi and M.~Drees,
  Phys.\ Lett.\  B {\bf 484} (2000),  183.

\bibitem{Kakizaki:2005en}
  M.~Kakizaki, S.~Matsumoto, Y.~Sato and M.~Senami,
  Phys.\ Rev.\  D {\bf 71}  (2005), 123522.

\bibitem{Burnell:2005hm}
  F.~Burnell and G.~D.~Kribs,
  Phys.\ Rev.\  D {\bf 73}  (2006), 015001.

\bibitem{Kong:2005hn}
  K.~Kong and K.~T.~Matchev,
  JHEP {\bf 0601}  (2006),  038.

\bibitem{Kakizaki:2006dz}
  M.~Kakizaki, S.~Matsumoto and M.~Senami,
  Phys.\ Rev.\  D {\bf 74}  (2006), 023504.

\bibitem{Agashe:2001xt}
  K.~Agashe, N.~G.~Deshpande and G.~H.~Wu,
  Phys.\ Lett.\  B {\bf 514}  (2001), 309.

\bibitem{Buras:2003mk}
  A.~J.~Buras, A.~Poschenrieder, M.~Spranger and A.~Weiler,
  Nucl.\ Phys.\  B {\bf 678}  (2004), 455.

\bibitem{Haisch:2007vb}
  U.~Haisch and A.~Weiler,
  Phys.\ Rev.\  D {\bf 76}  (2007), 034014.

\bibitem{Bashiry:2008en}
  V.~Bashiry and K.~Zeynali,
  Phys.\ Rev.\  D {\bf 79}  (2009), 033006.

\bibitem{Chakraverty:2002qk}
  D.~Chakraverty, K.~Huitu and A.~Kundu,
  Phys.\ Lett.\  B {\bf 558} (2003), 173.

\bibitem{Gogoladze:2006br}
  I.~Gogoladze and C.~Macesanu,
  Phys.\ Rev.\  D {\bf 74}  (2006), 093012.

\bibitem{Oikonomou:2006mh}
  V.~K.~Oikonomou, J.~D.~Vergados, C.~.C.~Moustakidis,
  Nucl.\ Phys.\  {\bf B773} (2007), 19-42.

\bibitem{Belanger:2010yx}
  G.~Belanger, M.~Kakizaki and A.~Pukhov,
  arXiv:1012.2577 [hep-ph].

\bibitem{Belanger:2006is}
  G.~Belanger, F.~Boudjema, A.~Pukhov and A.~Semenov,
  Comput.\ Phys.\ Commun.\  {\bf 176} (2007), 367.


\end{thebibliography}
\end{document}